\documentclass{emulateapj}

\slugcomment{Submitted to ApJ November 29, 2007 (revised March 20, 2008), MS ID 73647}

\shorttitle{Early injection and efficient mixing of $^{60}{\rm Fe}$}
\shortauthors{Dauphas et al.}

\begin{document}


\title{Iron-60 evidence for early injection and efficient mixing of stellar debris in the protosolar nebula}


\author{N. Dauphas\altaffilmark{1}, D.L. Cook\altaffilmark{2}, A. Sacarabany\altaffilmark{1}, C. Fr\"ohlich\altaffilmark{3}, A.M. Davis\altaffilmark{1}, M. Wadhwa\altaffilmark{4}, A. Pourmand\altaffilmark{1}, T. Rauscher\altaffilmark{5}, R. Gallino\altaffilmark{6,7}}

\altaffiltext{1}{Origins Laboratory, Department of the Geophysical Sciences and Enrico Fermi Institute, The University of Chicago, Chicago, IL 60637, USA; dauphas@uchicago.edu.}
\altaffiltext{2}{Department of Chemistry and Chemical Biology, Rutgers University, Piscataway, NJ 08854-8087, USA.}
\altaffiltext{3}{Department of Astronomy and Astrophysics, Enrico Fermi Institute, The University of Chicago, Chicago, IL 60637, USA.}
\altaffiltext{4}{School of Earth and Space Exploration, Arizona State University, Tempe, AZ 85287, USA.}
\altaffiltext{5}{Departement f\"ur Physik, Universit\"at Basel, CH-4056 Basel, Switzerland.}
\altaffiltext{6}{Dipartimento di Fisica Generale dell'Universita' di Torino, 10125 Torino, Italy.}
\altaffiltext{7}{Center for Stellar and Planetary Astrophysics, School of Mathematical Sciences, Monash University, Victoria 3800, Australia.}

\begin{abstract}

 Among extinct radioactivities present in meteorites, ${\rm ^{60}Fe}$ ($t_{1/2}=1.49$ Myr) plays a key role as a high-resolution chronometer, a heat source in planetesimals, and a fingerprint of the astrophysical setting of solar system formation. A critical issue with ${\rm ^{60}Fe}$ is that it could have been heterogeneously distributed in the protoplanetary disk, calling into question the efficiency of mixing in the solar nebula or the timing of ${\rm ^{60}Fe}$ injection relative to planetesimal formation. If this were the case, one would expect meteorites that did not incorporate ${\rm ^{60}Fe}$ (either because of late injection or incomplete mixing) to show ${\rm ^{60}Ni}$ deficits (from lack of ${\rm ^{60}Fe}$ decay) and collateral effects on other neutron-rich isotopes of Fe and Ni (coproduced with ${\rm ^{60}Fe}$ in core-collapse supernovae and AGB-stars). Here, we show that measured iron meteorites and chondrites have Fe and Ni isotopic compositions identical to Earth. This demonstrates that ${\rm ^{60}Fe}$ must have been injected into the protosolar nebula and mixed to less than 10 \% heterogeneity before formation of planetary bodies.

\end{abstract}





\section{Introduction}

Evidence for the presence of live ${\rm ^{60}Fe}$ has been found in meteoritic materials in the form of its decay product ${\rm ^{60}Ni}$ \citep{birck88,shukolyukov93a}. Recent {\it in situ} analyses of minerals from chondrites, with high Fe/Ni ratios, have improved our knowledge of ${\rm ^{60}Fe}$ abundance in the early solar system (ESS) \citep{tachibana03,mostefaoui05,tachibana06,guan07}. In particular, chondrule pyroxenes show ${\rm ^{60}Ni}$ excesses corresponding to ${\rm ^{60}Fe/^{56}Fe}$ ratios of 2.2 to $3.7\times 10^{-7}$ at the time of crystallization \citep{tachibana06}. This translates into an initial ratio of $5$ to $10\times 10^{-7}$ at the time of condensation of the first solids, calcium-aluminum-rich inclusions (CAIs), in the nebula \citep[assuming a time interval of 1.5 to 2.0 Myr between formation of chondrules and CAIs,][]{amelin02,kita05}. A critical question that remains to be answered is whether or not ${\rm ^{60}Fe}$ was homogeneously distributed in the ESS. Iron meteorites can help us address this issue because they may sample a different portion of the protosolar nebula than that sampled by chondrites \citep{bottke06}.  In addition, ${\rm ^{182}Hf}$ ($t_{1/2}=8.9$ Myr)-${\rm ^{182}W}$ chronology indicates that the parent bodies of iron meteorites accreted within $\sim 2$ Myr of formation of CAIs \citep{markowski06,schersten06,qin07}, when ${\rm ^{60}Fe}$ would have been extant if it had been present.  Several groups have analyzed Ni in iron meteorites and have found normal (terrestrial) isotopic compositions, within uncertainties \citep{cook06,quitte06,chen07,moynier07}. However, with improved precision, \citet{bizzarro07} reported deficits of 25 parts per million in $^{60}{\rm Ni}$ in several classes of early-formed differentiated meteorites (including iron meteorites); these deficits were not correlated with Fe/Ni ratios. They attributed the uniform deficits in $^{60}{\rm Ni}$ to the accretion of the parent bodies of these differentiated meteorites prior to the injection of ${\rm ^{60}Fe}$ into the ESS. 

In order to address the question of ${\rm ^{60}Fe}$ distribution in the solar nebula, we have analyzed Fe and Ni isotopic compositions of meteoritic metal. The paper is organized as follows. In \S 2, we describe the analytical methods used for measuring the isotopic compositions of Fe and Ni (Table 1).  In \S 3, we show how inefficient mixing or delayed injection of ${\rm ^{60}Fe}$ in the solar nebula should be associated with ${\rm ^{60}Ni}$ deficits (from lack of ${\rm ^{60}Fe}$ decay) and collateral anomalies in the neutron-rich isotopes ${\rm ^{58}Fe}$ and ${\rm ^{64}Ni}$ (coproduced with ${\rm ^{60}Fe}$ in core-collapse supernovae and AGB-stars by neutron-capture reactions). Comparisons between predictions and measurements limit the heterogeneity in the distribution of ${\rm ^{60}Fe}$. In \S 4, we set an upper-limit on the abundance of ${\rm ^{60}Fe}$ at the time of metal-silicate differentiation in planetesimals.  Finally, we discuss and summarize the implications of our results in \S 5.

\section{Materials and Methods}

For Fe analyses, we used aliquots of the samples dissolved by \citet{cook06}. The protocol used for separation of Fe for the search of isotopic anomalies is described in \citet{dauphas04,dauphas06}. The solutions of purified Fe were analyzed on a Neptune (Thermo Electron) multicollector inductively coupled plasma mass spectrometer (MC-ICPMS) at the University of Chicago using Al cones. Isotopes ${\rm ^{54}Fe}$, ${\rm ^{56}Fe}$, ${\rm ^{57}Fe}$, and ${\rm ^{58}Fe}$ were analyzed on Faraday collectors L2, Ax, H1, and H2. Isobaric interferences from ${\rm ^{54}Cr}$ and ${\rm ^{58}Ni}$ were corrected for by monitoring ${\rm ^{53}Cr}$ and ${\rm ^{60}Ni}$ on L4 and H4. The corrections from ${\rm ^{58}Ni}$ on ${\rm ^{58}Fe}$ ranged from $\sim$0 to 60 $\epsilon$. All collectors except Ax and H4 were connected to $10^{11}$ $\Omega$ amplifiers. A $10^{10}$ $\Omega$ amplifier was used for ${\rm ^{56}Fe}$ to prevent saturation and measure the minor isotopes at relatively high ion intensities. A $10^{12}$ $\Omega$ amplifier was used for ${\rm ^{60}Ni}$ in order to reduce the Johnson noise by a factor of $\sim2$ \citep{wieser05} and improve the precision of the Ni interference correction on ${\rm ^{58}Fe}$. The purified Fe solutions (10 ppm in  0.3 M HNO$_3$) were introduced into the MC-ICPMS using an Apex-Q+Spiro inlet system operated with Ar alone and a 100 $\mu$L/min PFA nebulizer. The measurements were made in High Resolution mode on peak shoulders \citep{weyer03}, allowing us to resolve molecular interferences from ${\rm ^{40}Ar^{14}N}$, ${\rm ^{40}Ar^{16}O}$, ${\rm ^{40}Ar^{16}O^{1}H}$, and ${\rm ^{40}Ar^{18}O}$ on Fe isotopes. The resulting ${\rm ^{56}Fe}$ ion intensity was 0.5-1.5 nA. On peak zero (OPZ) intensities from a blank solution were subtracted from all measurements. The data were acquired in a sequence of 25 cycles of 8.369 s each. Sample analyses were bracketed by measurements of the reference material IRMM-014 \citep{taylor92} run under identical conditions. A 90 s washout time was allowed between samples and standards. Instrumental and natural mass fractionation were corrected for by fixing the ${\rm ^{57}Fe/^{54}Fe}$ ratio to a constant value using the exponential law \citep{marechal99}. Each sample was analyzed between 12 and 24 times, and the average compositions are reported in $\epsilon$ units (Table \ref{table1}), $\epsilon ^i{\rm Fe}=\left[\left(^i{\rm Fe}/^{54}{\rm Fe}\right)_{\rm sample}/\left(^i{\rm Fe}/^{54}{\rm Fe}\right)_{\rm IRMM-014}-1\right]\times 10^4$.

For Ni, we used purified solutions already analyzed by \citet{cook06}. The conditions were identical to those used for Fe except for the concentration of the analyte (5 ppm), the cone material (Ni), and the gas injected in the Apex (Ar+N$_2$). Note that despite the use of Ni cones, the ion intensities measured in a blank solution before introducing any Ni into the mass spectrometer remain negligible (sample/OPZ=$2\times 10^4$). Isotopes ${\rm ^{58}Ni}$, ${\rm ^{60}Ni}$, ${\rm ^{61}Ni}$, ${\rm ^{62}Ni}$, and ${\rm ^{64}Ni}$ were analyzed on L2, L1, Ax, H1, and H2 Faraday collectors. Isobaric interferences from ${\rm ^{58}Fe}$ and ${\rm ^{64}Zn}$ were corrected for by monitoring ${\rm ^{57}Fe}$ and ${\rm ^{66}Zn}$ on L4 and H4 collectors. The corrections from ${\rm ^{64}Zn}$ on ${\rm ^{64}Ni}$ ranged from $\sim$ 5 to 20 $\epsilon$. Isotopes ${\rm ^{58}Ni}$ and ${\rm ^{66}Zn}$ were analyzed with $10^{10}$ and $10^{12}$ $\Omega$ amplifiers, respectively. High resolution was used to resolve interferences from ${\rm ^{40}Ar^{16}O^1H}$ and ${\rm ^{40}Ar^{18}O}$. Zoom optics were used to accommodate the mass dispersion (mass 57 to 66) in the collector array. OPZ intensities from blank solutions were subtracted from all measurements. The reference material used for sample-standard bracketing was SRM-986 \citep{gramlich89}. Each sample was analyzed 8 to 33 times, and the average compositions are reported in $\epsilon$ units, $\epsilon ^i{\rm Ni}=\left[\left(^i{\rm Ni}/^{58}{\rm Ni}\right)_{\rm sample}/\left(^i{\rm Ni}/^{58}{\rm Ni}\right)_{\rm SRM-986}-1\right]\times 10^4$ (the ratios have been corrected for mass-dependent fractionation by internal normalization). Two internal normalization schemes coexist for correcting mass-dependent fractionation. In \citet{cook06,quitte06,chen07}, the ${\rm ^{62}Ni/^{58}Ni}$ ratio was used while in \citet{birck88,bizzarro07}, the ${\rm ^{61}Ni/^{58}Ni}$ ratio was adopted. In Table \ref{table1}, we report $\epsilon$ values using both normalizations, allowing direct comparison with published data. Our measurements represent an improvement by a factor of 2 to 3 relative to the most precise analyses of \citet{cook06} made on the same solutions using a different instrument. As shown in Fig. \ref{fig:compbiz}, we did not detect any of the isotopic anomalies reported in \citet{bizzarro07}. The $\epsilon ^{62}{\rm Ni}$ anomalies reported in \citet{bizzarro07} are twice as large as the $\epsilon ^{60}{\rm Ni}$ values, entirely consistent with mass fractionation and suggestive of an isobaric interference on ${\rm ^{61}Ni}$.

\section{Early injection and efficient mixing of ${\rm ^{60}Fe}$}

Contrary to other extinct radioactivities, which could have been produced by particle irradiation within the solar system \citep[e.g., ${\rm ^{10}Be}$, ${\rm ^{26}Al}$,  ${\rm ^{41}Ca}$,  ${\rm ^{53}Mn}$,][]{mckeegan00,goswami01,gounelle01,leya03} or inherited from steady-state abundances in the interstellar medium \citep[e.g., ${\rm ^{92}Nb}$, ${\rm ^{129}I}$, ${\rm ^{146}Sm}$,  ${\rm ^{244}Pu}$,][]{clayton85,meyer00,dauphas03,wasserburg06}, ${\rm ^{60}Fe}$ requires injection from a nearby star such as a core-collapse supernova (cc-SN) or asymptotic giant branch star (AGB) shortly before formation of the solar system \citep{cameron77,wasserburg98,meyer00,mckeegan03,wasserburg06}. Interestingly, a cc-SN explosion or passing AGB-star could have triggered the collapse of the molecular cloud core that made our solar system \citep{cameron77,cameron93,foster96,vanhala00}. The probability of encounters between AGB stars and molecular clouds is small, whereas cc-SNs are associated with star-forming regions. Thus, cc-SNs are more likely candidates for explaining the presence of ${\rm ^{60}Fe}$ in meteorites than AGBs \citep{kastner94}. An alternative to the triggered molecular cloud core collapse hypothesis is direct injection of ${\rm ^{60}Fe}$ into the protoplanetary disk \citep{chevalier00,hester05,ouellette07}. The likelihood of such an injection is $<1$ \% \citep{williams07,gounelle08}.

\subsection{${\rm ^{60}Ni}$ deficits?}

 All Ni measurements have been corrected for mass-dependent fractionation by internal normalization to the terrestrial ${\rm ^{62}Ni/^{58}Ni}$ ratio. Accordingly, terrestrial rocks related by the laws of mass-dependent fractionation should have $\epsilon =0$ for the ${\rm ^{60}Ni/^{58}Ni}$, ${\rm ^{61}Ni/^{58}Ni}$, and ${\rm ^{64}Ni/^{58}Ni}$ ratios. As shown in Fig. \ref{fig:compilcomp}, we found normal (terrestrial) ${\rm ^{60}Ni}$ isotopic compositions in all analyzed meteorites (weighted average $\epsilon ^{60}{\rm Ni}=0.006\pm 0.011$). This conclusion agrees with previous studies \citep{cook06,quitte06,chen07,moynier07} but with significantly improved precision. The discrepancy with \citet{bizzarro07} is not due to differences in sampling \citep[meteorites of the same type were analyzed in][and this study]{bizzarro07} or the choice of the ratio used for correcting mass-dependent fractionation (Fig. \ref{fig:compbiz}). If ${\rm ^{60}Fe}$ were heterogeneously distributed in the solar nebula, one would expect to see $\epsilon ^{60}{\rm Ni}$ deficits in a hypothetical reservoir free of ${\rm ^{60}Fe}$ relative to the CHondritic Uniform Reservoir (CHUR, see Appendix for details),
\begin{equation}
\epsilon ^{60}{\rm Ni}_0=-Q\left(\frac{\rm ^{60}Fe}{\rm ^{56}Fe}\right)_{t_0,{\rm CHUR}},
\label{eq:l1}
\end{equation}
where $Q=\left(^{56}{\rm Fe}/^{60}{\rm Ni}\right)_{\rm CHUR}\times 10^4=61.35\times 10^4$ \citep{lodders03} and $t_0$ is the time of CAI formation. Here it is assumed that the terrestrial reference material has the same Ni isotopic composition as CHUR ($\rm ^{60}Fe$ was present in the two reservoirs at a similar level). For $\left(^{60}{\rm Fe}/^{56}{\rm Fe}\right)_{t_0,{\rm CHUR}}=5\times 10^{-7}$ \citep{tachibana06}, the predicted effect is $\epsilon ^{60}{\rm Ni}=-0.31$, nearly the same as the value reported in differentiated meteorites by \citet{bizzarro07}. Our high-precision measurements rule out the presence of such a deficit (Figs. \ref{fig:compbiz}, \ref{fig:compilcomp}).  The isotopic composition of a hypothetical reservoir containing ${\rm ^{60}Fe}$ at a lower or higher level than CHUR (denoted with {\it r} subscript) is given by $\epsilon ^{60}{\rm Ni}_{r}=Q\left[\left(^{60}{\rm Fe}/^{56}{\rm Fe}\right)_{t_0,r}-\left(^{60}{\rm Fe}/^{56}{\rm Fe}\right)_{t_0,{\rm CHUR}}\right]$ (see Appendix, Eq. \ref{eq:het}). Using $\left(^{60}{\rm Fe}/^{56}{\rm Fe}\right)_{t_0,{\rm CHUR}}=5\times 10^{-7}$, a reservoir with an initial ratio of $>5.8\times 10^{-7}$, would have $\epsilon ^{60}{\rm Ni}_r>+0.05$. Similarly, a reservoir with an initial ratio of $<4.2\times 10^{-7}$, would  have $\epsilon ^{60}{\rm Ni}_r<-0.05$. Thus, the $\pm 0.05$ dispersion in $\epsilon ^{60}{\rm Ni}$ around zero only allows for a heterogeneity of less than $\sim 15$ \% in the initial $\left(^{60}{\rm Fe}/^{56}{\rm Fe}\right)_{t_0}$ ratio (if the ratio is $5\times 10^{-7}$; a higher initial value would allow even less heterogeneity). This observation is consistent with the degree of homogeneity inferred for ${\rm ^{26}Al}$ in the accretion region of meteorite parent bodies \citep{mckeegan03} as well as results of dynamical modeling, which show that passive tracers are mixed in the nebula at the 10 \% level on time-scales of a few thousand years \citep{boss07}.

\subsection{Collateral nucleosynthetic effects on ${\rm ^{58}Fe}$ and ${\rm ^{64}Ni}$?}

A powerful means of investigating whether ${\rm ^{60}Fe}$ was heterogeneously distributed in the ESS is to search for collateral anomalies in other isotopes of Fe \citep{nichols99,sahijpal06,gounelle07}. Solar system Fe was produced by different nucleosynthetic processes (nuclear statistical equilibrium and neutron capture) in different stars \citep[$\sim 1/2$ comes from cc-SN and the rest from SNIa,][]{iwamoto99}. Its isotopic composition represents the integrated contribution over $\sim9$ Gyr of SNIa and cc-SN of different metallicities and initial masses. Thus, Fe isotopes are unlikely to be present in exact solar proportions in the cc-SN or AGB star that injected $^{60}{\rm Fe}$ into the nascent solar system. Addition of even minute amounts of stellar ejecta containing ${\rm ^{60}Fe}$ should leave a diagnostic fingerprint on non-radiogenic Fe isotopes if it is missing from other parts of the solar system \citep[as is suggested by][]{sugiura06,bizzarro07,quitte07}. The virtue of this approach is that no chemical fractionation can decouple ${\rm ^{60}Fe}$ from stable Fe isotopes \citep[whether ${\rm ^{60}Fe}$ is injected in the form of dust or gas,][is inconsequential]{ouellette07}.  

The bulk ejecta of a cc-SN could not have been injected into the ESS because such a scenario would overproduce ${\rm ^{41}Ca}$, ${\rm ^{53}Mn}$, and ${\rm ^{60}Fe}$ relative to ${\rm ^{26}Al}$ \citep{meyer00}. A possible solution to this problem is to invoke an injection mass cut different from the mass cut of the remnant \citep{cameron95,meyer00}. The meteoritic abundances of ${\rm ^{26}Al}$, ${\rm ^{41}Ca}$, and ${\rm ^{60}Fe}$ can be explained with a 25 M$_{\odot}$ (solar mass) cc-SN using a free decay interval of $\sim 1$ Myr and an injection mass cut of $\sim 7$ M$_{\odot}$ that corresponds to the edge of the He exhausted core \citep{meyer00}. In a cc-SN, Fe isotopes are produced in different layers by different nuclear reactions. Iron-54, 56, and 57 are produced as radioactive progenitors (e.g., ${\rm ^{56}Ni}$) in the inner region of the cc-SN by nuclear statistical equilibrium associated with explosive Si-burning. This process results in synthesis of isotopes with approximately equal numbers of neutrons and protons. For this reason, ${\rm ^{58}Fe}$ and ${\rm ^{60}Fe}$, which have large excesses of neutrons, are not produced in appreciable quantities. These two isotopes are synthesized in more external regions by neutron capture reactions on preexisting Fe isotopes (e.g., Fig. \ref{fig:tommy}). Using an injection mass cut similar to that used in \citet{meyer00}, one would predict that ${\rm ^{56}Fe/^{54}Fe}$ and ${\rm ^{57}Fe/^{54}Fe}$ ratios should be similar to solar values because the inventory of these isotopes in the ejecta is dominated by unprocessed material from the envelope. However, one would also predict that the ${\rm ^{58}Fe/^{54}Fe}$ ratio should be higher than solar because ${\rm ^{60}Fe}$ is accompanied by ${\rm ^{58}Fe}$. Therefore, any portion of the solar system deficient in ${\rm ^{60}Fe}$ should also exhibit negative anomalies in $\epsilon ^{58}{\rm Fe}$. The ${\rm ^{58}Fe}$ isotopic composition of a reservoir missing the cc-SN or AGB component relative to CHUR is (in $\epsilon$ with internal normalization, although in this case the isotopic variations are nucleosynthetic in origin and are not due to mass fractionation),
\begin{equation}
\epsilon ^{58}{\rm Fe}_0=-\frac{\rho ^{58}_{\rm Fe}-\mu ^{58}_{\rm Fe}\,\rho ^{57}_{\rm Fe}}{1+\rho ^{56}_{\rm Fe}}\,
\frac{\left(^{60}{\rm Fe}/^{56}{\rm Fe}\right)_{t_0,{\rm CHUR}}}{\left(^{60}{\rm Fe}/^{56}{\rm Fe}\right)_{\rm ccSN/AGB}}e^{\lambda_{60}\Delta t}\times 10^4,
\label{eq:l2}
\end{equation}
where $\rho ^{58}_{\rm Fe}=\left(^{58}{\rm Fe}/^{54}{\rm Fe}\right)_{\rm ccSN/AGB}/\left(^{58}{\rm Fe}/^{54}{\rm Fe}\right)_{\rm CHUR}-1$, $\mu ^{58}_{\rm Fe}=(58-54)/(57-54)$, $\lambda _{60}$ is the decay constant of ${\rm ^{60}Fe}$, and $\Delta t$ is the free decay interval between production in the cc-SN or AGB and injection into the solar system \citep[the derivation of the general equation is given in Appendix, also see][]{dauphas04b}. If one allows for $\Delta t>0$, the sizes of the collateral effects increase because ${\rm ^{60}Fe}$ has time to decay before being injected and the contribution of material from the cc-SN or AGB must increase in order to explain $\left(^{60}{\rm Fe}/^{56}{\rm Fe}\right)_{t_0,{\rm CHUR}}$, which is fixed. In this study, we use $\Delta t=0$ and present collateral isotopic effects that must therefore be treated as lower limits. In Fig. \ref{fig:modelres}A, we have computed $\epsilon ^{58}{\rm Fe}$ deficits using yields from 15, 19, and 25 M$_{\odot}$ cc-SN models as a function of injection mass cut \citep{rauscher02}. Note that the ${\rm ^{60}Fe}$ yields in these models are thought to be overestimated \citep{limongi06,woosley07} and the collateral effects on $\epsilon ^{58}{\rm Fe}$ may be larger than what is calculated here (the ${\rm ^{60}Fe/^{56}Fe}$ production ratio appears in the denominator of Eq. 2). However, the combined uncertainties in the neutron capture reactions governing the synthesis of ${\rm ^{60}Fe}$ are not expected to reduce the production of ${\rm ^{60}Fe}$ by more than a factor of two \citep{woosley07}. The predicted deficits on $\epsilon ^{58}{\rm Fe}$ are 3 epsilon units or more, depending on the injection mass cut. Such effects are inconsistent with $\epsilon ^{58}{\rm Fe}$ measured in iron meteorites, which are all within $0.3$ of the terrestrial and chondritic value \citep[Table \ref{table1}, also see][]{dauphas04}. This, together with normal $\epsilon ^{60}{\rm Ni}$, provides compelling evidence against delayed injection or incomplete mixing of ${\rm ^{60}Fe}$ in the solar nebula.  The dispersion of $\pm 0.3$ in $\epsilon ^{58}{\rm Fe}$ only allows for a heterogeneity of less than $\sim 10$ \% in the initial $\left(^{60}{\rm Fe}/^{56}{\rm Fe}\right)_{t_0}$ ratio (if the ratio is $5\times 10^{-7}$; see Appendix Eq. \ref{eq:het2}).

We can also quantify the degree of heterogeneity of ${\rm ^{60}Fe}$ in the ESS by measuring collateral effects in the nonradiogenic isotopes of Ni. If ${\rm ^{60}Fe}$ was missing from parts of the solar system, then one would expect such reservoirs to also show deficits in ${\rm ^{64}Ni}$, the most neutron-rich isotope of nickel. A complication arises with collateral anomalies in Ni isotopes because one must make the assumption that Fe and Ni from the cc-SN are not decoupled during injection into the ESS. In the 15, 19, and 25 M$_{\odot}$ cc-SN models, ${\rm ^{60}Fe}$ is produced in layers where ${\rm C/O<1}$ and Fe would presumably condense as metal \citep{lattimer78}. Under those circumstances, Ni is unlikely to be decoupled from Fe. The predicted collateral ${\rm ^{64}Ni}$ isotope effects are given by, 
\begin{equation}
\epsilon ^{64}{\rm Ni}_0=-c\,\frac{\rho ^{64}_{\rm Ni}-\mu ^{64}_{\rm Ni}\,\rho ^{62}_{\rm Ni}}{1+\rho ^{56}_{\rm Fe}}\,
\frac{\left(^{60}{\rm Fe}/^{56}{\rm Fe}\right)_{t_0,{\rm CHUR}}}{\left(^{60}{\rm Fe}/^{56}{\rm Fe}\right)_{\rm ccSN/AGB}}e^{\lambda_{60}\Delta t}\times 10^4 ,
\label{eq:l3}
\end{equation}
where $\rho ^{64}_{\rm Ni}=\left(^{64}{\rm Ni}/^{58}{\rm Ni}\right)_{\rm ccSN/AGB}/\left(^{64}{\rm Ni}/^{58}{\rm Ni}\right)_{\rm CHUR}-1$, $\mu ^{64}_{\rm Ni}=(64-58)/(62-58)$, and $c=\left(^{58}{\rm Ni}/^{54}{\rm Fe}\right)_{\rm ccSN/AGB}/\left(^{58}{\rm Ni}/^{54}{\rm Fe}\right)_{\rm CHUR}$ \citep[the derivation of the general equation is given in Appendix, also see][]{dauphas04b}. If some parts of the solar system did not incorporate ${\rm ^{60}Fe}$ from a cc-SN, then one would predict $\epsilon ^{64}{\rm Ni}$ of $-2$ or lower (Fig. \ref{fig:modelres}B). This contradicts measurements, which show normal $\epsilon ^{64}{\rm Ni}$ within 0.2. This limits the possible heterogeneity of the initial $\left({\rm ^{60}Fe/^{56}Fe}\right)_{t_0}$ ratio to within 10 \% (if the value is $5\times 10^{-7}$; see Appendix Eq. \ref{eq:het3}).

Although less likely than cc-SN \citep{kastner94,williams07}, AGB-stars represent an alternative source for ${\rm ^{26}Al}$  and ${\rm ^{60}Fe}$ in the ESS \citep{cameron93,wasserburg98,wasserburg06}. In AGBs, neutrons produced by the ${\rm ^{22}Ne(\alpha ,n)^{25}Mg}$ reaction can be captured on ${\rm ^{54}Fe}$, ${\rm ^{56}Fe}$, and ${\rm ^{57}Fe}$ to produce ${\rm ^{58}Fe}$ and ${\rm ^{60}Fe}$. The predicted $\epsilon ^{58}{\rm Fe}$ deficit (Eq. 2) for a 3 M$_{\odot}$ initial mass star with solar metallicity and standard ${\rm ^{13}C}$-pocket \citep{wasserburg06} is $\sim -133$ (for an initial $^{60}{\rm Fe}/^{56}{\rm Fe}=5\times 10^{-7}$). Again, no such deficits were found ($\epsilon ^{58}{\rm Fe}=0\pm 0.3$). Similarly, the predicted $\epsilon ^{64}{\rm Ni}$ deficit of $-125$ is not observed in meteorites ($\epsilon ^{64}{\rm Ni}=0\pm 0.2$). Thus, if the source of ${\rm ^{60}Fe}$ in the ESS was a nearby AGB-star, limits on the possible heterogeneous distribution of ${\rm ^{60}Fe}$ in the protosolar nebula are even more stringent ($\pm 0.2$ \%) than in the case of a cc-SN.  

Analyses of Fe and Ni stable isotopic compositions of meteorites limit the heterogeneity in the distribution of ${\rm ^{60}Fe}$ to less than $\sim 10$ \%. This is consistent with contemporaneous injection of ${\rm ^{60}Fe}$ and ${\rm ^{26}Al}$ followed by efficient mixing in the protosolar disk \citep{boss07}. It is worth noting that the scale relevant to this study is that of planetesimals. Stable isotopic anomalies produced by neutron-rich nuclear statistical equilibrium \citep{hartmann85} have been detected in CAIs for Fe and Ni \citep{birck88,volkening89} and one cannot exclude that ${\rm ^{60}Fe}$ was heterogeneously distributed at such scale (subcentimeter objects).  

\section{Upper limit on ${\rm ^{60}Fe}$ abundance}

Having established that ${\rm ^{60}Fe}$ was homogeneously distributed in the ESS, we can now turn to estimating its abundance. Evidence from ${\rm ^{182}Hf-^{182}W}$ systematics indicates that separation between metal and silicate occurred within ~2 Myr of condensation of the first solids in the nebula \citep{markowski06,schersten06,qin07}. In Fig. \ref{fig:isochron}, we plot $\epsilon ^{60}{\rm Ni}$ (using ${\rm ^{62}Ni/^{58}Ni}$ normalization) as a function of the Fe/Ni fractionation factor for several types of samples. For Bishunpur metal, we use the Fe/Ni ratio measured by \citet{cook06}; $f_{\rm Fe/Ni}={\rm (Fe/Ni)/(Fe/Ni)_{\rm CHUR}-1}=-0.221$. Part of the fractionation between Fe and Ni in iron meteorites was produced during fractional crystallization of the metallic core, which most likely took place after ${\rm ^{60}Fe}$ had decayed. Under those circumstances, the relevant Fe/Ni fractionation factor is that of the parental melt, established when metallic cores segregated from silicate mantles. We therefore plot the average $\epsilon ^{60}{\rm Ni}$ for each magmatic iron meteorite group (Table \ref{table1}) as a function of the fractionation factor of the parent melt \citep[$f_{\rm Fe/Ni}=-0.094$ and $-0.329$ for IIAB and IIIAB, respectively,][]{jones83}. If ${\rm ^{60}Fe}$ was present when metal segregated from silicate (taken to represent a single event precisely defined in time), there should be a linear correlation between $\epsilon ^{60}{\rm Ni}$ and the Fe/Ni fractionation factor relative to chondrites \citep[][see Appendix for details]{jacobsen84},
\begin{equation}
\epsilon ^{60}{\rm Ni}=Q\left(\frac{\rm ^{60}Fe}{\rm ^{56}Fe}\right)_{t_i}\,f_{\rm Fe/Ni},
\label{eq:l4}
\end{equation}
where $Q$ has the same definition as in Eq. \ref{eq:l1} and $t_i$ is the time of Fe/Ni fractionation. Bishunpur metal and IIIAB irons provide the most stringent constraints on the ${\rm ^{60}Fe/^{56}Fe}$ ratio at the time of metal-silicate differentiation, which is conservatively estimated to be $<2\times 10^{-7}$ in both cases. This value is in agreement with the ${\rm ^{60}Fe/^{56}Fe}$ ratio measured in Bishunpur by {\it in situ} techniques \citep[$1.08\pm 0.23\times 10^{-7}$ in sulfides and $1.9\pm 1.1\times 10^{-7}$ in chondrules]{tachibana03,tachibana07} and the value inferred from pallasite and magmatic iron meteorite measurements \citep[marginally different from 0, $\sim 1$ to $8\times 10^{-7}$]{cook06}.

The ${\rm ^{182}Hf-^{182}W}$ date of core formation in the parent-body of IIIAB iron meteorites is $<1.6$ Myr after CAI formation \citep{markowski06,schersten06,qin07}. No precise ${\rm ^{182}Hf-^{182}W}$ age of metal in LL chondrites is available \citep{yin02} but \citet{kleine08} showed that in primitive H chondrites, metal-silicate differentiation occurred $1.7\pm 0.7$ Myr after CAI and was coeval with chondrule formation \citep{amelin02,kita05}. If we tentatively correct for a possible delay of up to 2 Myr between CAI formation and metal-silicate differentiation, the $\epsilon ^{60}{\rm Ni}$ values measured in meteoritic metal translate into an upper limit of $<6\times 10^{-7}$ for the initial ${\rm ^{60}Fe/^{56}Fe}$ ratio at the time  of condensation of the first solids in the solar nebula. This result is consistent with the initial value of $5-10\times 10^{-7}$ derived from {\it in situ} isotopic analysis of chondrule pyroxenes \citep{tachibana06}.

\section{Summary and conclusion}

The presence of ${\rm ^{60}Fe}$ in meteorites can only be explained by injection into the solar nebula of debris from a nearby AGB or cc-SN. Homogeneity of short-lived isotopes in  the solar nebula is one of the key unsettled questions at the present time. In order to address this question, we have measured the isotopic compositions of Fe and Ni in meteoritic metal. The data have clear implications:

\begin{itemize}

\item If ${\rm ^{60}Fe}$ was not well mixed or if it was injected late into the solar nebula, one would expect the meteorites that did not incorporate ${\rm ^{60}Fe}$ to show ${\rm ^{60}Ni}$ deficits (from lack of ${\rm ^{60}Fe}$ decay). Our measurements show that within uncertainties, the isotopic abundance of ${\rm ^{60}Ni}$ in metal from IIAB, IIIAB, PMG, and LL meteorites is constant, consistent with homogeneous distribution of ${\rm ^{60}Fe}$ in the solar nebula (less than 15 \% heterogeneity in the ${\rm ^{60}Fe/^{56}Fe}$ ratio).

\item Iron-60 is synthesized in cc-SN and AGB stars by neutron-capture reactions. If some parts of the solar system did not incorporate ${\rm ^{60}Fe}$, they should show deficits in the abundances of other isotopes also produced by neutron-capture, ${\rm ^{58}Fe}$ and ${\rm ^{64}Ni}$. These isotopes have very low isotopic abundances and are difficult to analyze. Our study reveals that the abundances of ${\rm ^{58}Fe}$ and ${\rm ^{64}Ni}$ in meteorites are identical within uncertainties to abundances measured in terrestrial materials. Comparisons with model predictions limit the possible heterogeneous distribution of ${\rm ^{60}Fe}$ to less than 10 \%.

\item Metal from IIAB, IIIAB, and LL meteorites have fractionated Fe/Ni ratios relative to CHUR. If Fe/Ni fractionation occurred in the presence of ${\rm ^{60}Fe}$, one would expect to find isotopic variations in the abundance of ${\rm ^{60}Ni}$, possibly correlated with Fe/Ni ratios. No correlation was found, providing an upper-limit on the initial ${\rm ^{60}Fe/^{56}Fe}$ ratio in the ESS of around $<6\times 10^{-7}$. This upper-limit agrees with recent {\it in situ} isotopic analyses of chondrule pyroxenes, which give an initial ratio of $5$ to $10\times 10^{-7}$ \citep{tachibana06}.

\end{itemize}

The  degree of mixing of ${\rm ^{60}Fe}$ in the ESS is quantified for the first time in this study. We conclude that ${\rm ^{60}Fe}$ must have been homogeneously distributed (less than 10 \% heterogeneity in the ${\rm ^{60}Fe/^{56}Fe}$ ratio). This agrees with results of dynamical modeling \citep{boss07}, showing that passive tracers are mixed in a turbulent  nebula at the 10 \% level within several thousand years. A second important conclusion is that ${\rm ^{60}Fe}$ was injected before formation of planetary bodies. This supports the idea that the presolar molecular cloud core  was triggered into collapse by interaction with a nearby star \citep{cameron77,cameron93}. Iron-60 was homogeneously  distributed in the solar nebula and can therefore be used as a  reliable short-lived chronometer for ESS events. This is particularly important given the converging lines of evidence indicating that the key events that shaped our solar system took place shortly after condensation of the first  solids in the solar nebula. 

\acknowledgments

This inaugural paper for the MC-ICPMS facility at the University of Chicago is dedicated to the memory of Toshiko K. Mayeda. Support from B.P. Lynch, E.E. Hunter, D.B. Rowley, M.J. Foote and discussions with R.N. Clayton, L. Grossman, J.W. Truran, F.M. Richter, F.-Z. Teng, D.A. Papanastassiou, J.-L. Birck were greatly appreciated. An anonymous reviewer provided useful comments that improved the quality of the manuscript. This work was supported by a fellowship from the Packard Foundation to N.D., the France-Chicago Center (N.D.), NASA grants NNG06GG75G (N.D.), NNG06GF19G (A.M.D.), NNG05GG22G (M.W.), NNG06GF13G (to R.N. Clayton for support of David L. Cook), Swiss NSF grant 20-113984/1 (T.R.), and Italian MIUR-Cofin 2006 (R.G.). 



\appendix

\section{Derivation of Eq. 1}

Over time, all initial ${\rm ^{60}Fe}$ present in CHUR was converted into ${\rm ^{60}Ni}$,
\begin{equation}
^{60}{\rm Ni}_{\rm CHUR}=^{60}{\rm Ni}_0+^{60}{\rm Fe}_{t_0,{\rm CHUR}}.
\end{equation}
Introducing isotopic ratios into this equation,
\begin{equation}
\left(\frac{\rm ^{60}Ni}{\rm ^{58}Ni}\right)_{\rm CHUR}=\left(\frac{\rm ^{60}Ni}{\rm ^{58}Ni}\right)_{0}+\left(\frac{\rm ^{60}Fe}{\rm ^{56}Fe}\right)_{t_0,{\rm CHUR}}\left(\frac{\rm ^{56}Fe}{\rm ^{58}Ni}\right)_{\rm CHUR},
\label{eq:a2}
\end{equation}
where $\left(^{60}{\rm Ni}/^{58}{\rm Ni}\right)_0$ is also the Ni isotopic composition of any reservoir missing ${\rm ^{60}Fe}$ (either because of delayed injection of incomplete mixing). Dividing both sides by $\left(^{60}{\rm Ni}/^{58}{\rm Ni}\right)_{\rm CHUR}$ gives,
\begin{equation}
\frac{\left(^{60}{\rm Ni}/^{58}{\rm Ni}\right)_0}{\left(^{60}{\rm Ni}/^{58}{\rm Ni}\right)_{\rm CHUR}}=1-\left(\frac{\rm ^{56}Fe}{\rm ^{60}Ni}\right)_{\rm CHUR}\left(\frac{\rm ^{60}Fe}{\rm ^{56}Fe}\right)_{t_0,{\rm CHUR}}.
\end{equation}
Expressed in $\epsilon$ notation relative to CHUR, this translates into,
\begin{equation}
\epsilon ^{60}{\rm Ni}_0=-10^4 \left(\frac{\rm ^{56}Fe}{\rm ^{60}Ni}\right)_{\rm CHUR}\left(\frac{\rm ^{60}Fe}{\rm ^{56}Fe}\right)_{t_0,{\rm CHUR}}.
\end{equation}
Introducing $Q=\left(^{56}{\rm Fe}/^{60}{\rm Ni}\right)_{\rm CHUR}\times 10^4$, this takes the form (Eq. \ref{eq:l1}),
\begin{equation}
\epsilon ^{60}{\rm Ni}_0=-Q\left(\frac{\rm ^{60}Fe}{\rm ^{56}Fe}\right)_{t_0,{\rm CHUR}}.
\end{equation}
Let us now consider a hypothetical reservoir $r$ with an initial $\left({\rm ^{60}Fe/^{56}Fe}\right)_{t_0,r}$ ratio different from CHUR but with chondritic Fe/Ni ratio. An equation similar to Eq \ref{eq:a2} can be posed,
\begin{equation}
\left(\frac{\rm ^{60}Ni}{\rm ^{58}Ni}\right)_{r}=\left(\frac{\rm ^{60}Ni}{\rm ^{58}Ni}\right)_{0}+\left(\frac{\rm ^{60}Fe}{\rm ^{56}Fe}\right)_{t_0,r}\left(\frac{\rm ^{56}Fe}{\rm ^{58}Ni}\right)_{\rm CHUR}.
\end{equation}
Dividing  the 2 sides by ${\rm (^{60}Ni/^{58}Ni)_{CHUR}}$, it is easy to show that,
\begin{equation}
\epsilon ^{60}{\rm Ni}_{r}=\epsilon ^{60}{\rm Ni}_0+Q\left(^{60}{\rm Fe}/^{56}{\rm Fe}\right)_{t_0,r},
\end{equation}
which can also be written in the form,
\begin{equation}
\epsilon ^{60}{\rm Ni}_{r}=Q\left[\left(^{60}{\rm Fe}/^{56}{\rm Fe}\right)_{t_0,r}-\left(^{60}{\rm Fe}/^{56}{\rm Fe}\right)_{t_0,{\rm CHUR}}\right].
\label{eq:het}
\end{equation}
This equation can be used to quantify the degree of heterogeneity in ${\rm ^{60}Fe/^{56}Fe}$ ratios based on $\epsilon ^{60}{\rm Ni}$ measurements.

\section{Derivation of Eq. 2}

Let us denote $^{60}{\rm Fe}_{\rm inj}$ the amount of $^{60}{\rm Fe}$ injected into the ESS. If $\Delta t$ is the time that separates nucleosynthesis in AGB or cc-SN and injection into the ESS,
\begin{equation}
^{60}{\rm Fe}_{\rm inj}=^{60}{\rm Fe}_{\rm ccSN/AGB}\,e^{-\lambda _{60}\Delta t}.
\end{equation}
The amount of ${\rm ^{60}Fe}$ injected is equal to the amount of ${\rm ^{60}Fe}$ in chondrites (back-calculated at the time of condensation of the first solids in the protosolar nebula),
 \begin{equation}
\left(\frac{^{60}{\rm Fe}}{^{56}{\rm Fe}}\right)_{\rm inj}{^{56}{\rm Fe}_{\rm inj}}=\left(\frac{^{60}{\rm Fe}}{^{56}{\rm Fe}}\right)_{t_0,{\rm CHUR}}{^{56}{\rm Fe}_{\rm CHUR}}.
\end{equation}
If we define $x=^{56}{\rm Fe}_{\rm inj}/^{56}{\rm Fe}_{\rm CHUR}=^{56}{\rm Fe}_{\rm ccSN/AGB}/^{56}{\rm Fe}_{\rm CHUR}$, it follows from the previous equations that,
\begin{equation}
x=\frac{\left(^{60}{\rm Fe}/^{56}{\rm Fe}\right)_{t_0,{\rm CHUR}}}{\left(^{60}{\rm Fe}/^{56}{\rm Fe}\right)_{\rm ccSN/AGB}}e^{\lambda_{60}\Delta t}.
\label{eq:x1}
\end{equation}
Let us now define $y=^{54}{\rm Fe}_{\rm ccSN/AGB}/^{54}{\rm Fe}_{\rm CHUR}$,
\begin{equation}
y=\frac{\rm \left(^{54}Fe/^{56}Fe\right)_{\rm ccSN/AGB}}{\rm \left(^{54}Fe/^{56}Fe\right)_{\rm CHUR}}\frac{\rm ^{56}Fe_{ccSN/AGB}}{\rm ^{56}Fe_{CHUR}}.
\label{eq:y1}
\end{equation}
Using the notation,
\begin{equation}
\rho ^i_{\rm Fe}=\left(^i{\rm Fe}/^{54}{\rm Fe}\right)_{\rm ccSN/AGB}/\left(^i{\rm Fe}/^{54}{\rm Fe}\right)_{\rm CHUR}-1,
\end{equation}
Eq. \ref{eq:y1} can be rewritten in the form,
\begin{equation}
y=x/\left(1+\rho ^{56}_{\rm Fe}\right).
\label{eq:y2}
\end{equation}
For any isotopes $^i{\rm Fe}$, we can pose a conservation equation,
\begin{equation}
\left(\frac{^i{\rm Fe}}{^{54}{\rm Fe}}\right)_{\rm CHUR}{^{54}{\rm Fe}_{\rm CHUR}}=\left(\frac{^i{\rm Fe}}{^{54}{\rm Fe}}\right)_{\rm 0}{^{54}{\rm Fe}_{\rm 0}}+\left(\frac{^i{\rm Fe}}{^{54}{\rm Fe}}\right)_{\rm ccSN/AGB}{^{54}{\rm Fe}_{\rm ccSN/AGB}},
\end{equation}
where the $0$ subscript denotes the composition of CHUR prior to ${\rm ^{60}Fe}$ injection. It is also the composition of any planetary or meteoritic component that did not receive ${\rm ^{60}Fe}$ either because it formed before ${\rm ^{60}Fe}$ was injected into the ESS or it was accreted in a region of the nebula that did not incorporate ${\rm ^{60}Fe}$ due to incomplete mixing. Introducing $y$ and $\rho ^i_{\rm Fe}$ in this equation, it is straightforward to show that the isotopic composition of a reservoir missing the cc-SN or AGB component relative to CHUR is,
\begin{equation}
\frac{\left(^i{\rm Fe}/^{54}{\rm Fe}\right)_0}{\left(^i{\rm Fe}/^{54}{\rm Fe}\right)_{\rm CHUR}}=\frac{1-(1+\rho ^i_{\rm Fe})y}{1-y}.
\end{equation}
In $\epsilon ^{\ast}=\left({\rm R/R_{CHUR}}-1\right)\times 10^4$ notation, this takes the form,
\begin{equation}
\epsilon ^{i\ast}{\rm Fe}=-\frac{\rho ^i_{\rm Fe}y}{1-y}\times 10^4,
\label{eq:ep1}
\end{equation}
where the $\ast$ superscript indicates that the ratios have not been internally normalized. Eq. \ref{eq:ep1} can be applied to ${\rm ^{57}Fe}$,
\begin{equation}
\epsilon ^{57\ast}{\rm Fe}=-\frac{\rho ^{57}_{\rm Fe}y}{1-y}\times 10^4.
\label{eq:ep1b}
\end{equation}
 Isotopic analyses are corrected for mass fractionation by internal normalization to a constant ${\rm ^{57}Fe/^{54}Fe}$ ratio. Internally normalized $\epsilon$ values are computed using the following formula,
\begin{equation}
\epsilon ^i{\rm Fe}=\epsilon ^{i\ast}{\rm Fe}-\mu ^i_{\rm Fe}\epsilon ^{57\ast}{\rm Fe},
\end{equation}
where $\mu ^i_{\rm Fe}=(i-54)/(57-54)$. We therefore have,
\begin{equation}
\epsilon ^i{\rm Fe}_0=-\frac{y}{1-y}\left(\rho ^i_{\rm Fe}-\mu ^i_{\rm Fe}\rho ^{57}_{\rm Fe}\right)\times 10^4.
\label{eq:ep2}
\end{equation}
From Eq. \ref{eq:y2}, we have $y=x/\left(1+\rho ^{56}_{\rm Fe}\right)$ and thus $y/\left(1-y\right)=x/\left(1+\rho ^{56}_{\rm Fe}-x\right)$. The fraction of the solar system inventory of Fe contributed by nucleosynthesis in a nearby star must have been small ($x\ll 1+\rho ^{56}_{\rm Fe}$). Using $y/\left(1-y\right)\simeq x/\left(1+\rho ^{56}_{\rm Fe}\right)$ and the expression of $x$ given in Eq. \ref{eq:x1}, Eq. \ref{eq:ep2} takes the form (Eq. \ref{eq:l2}),
\begin{equation}
\epsilon ^i{\rm Fe}_0\simeq-\frac{\rho ^i_{\rm Fe}-\mu ^i_{\rm Fe}\,\rho ^{57}_{\rm Fe}}{1+\rho ^{56}_{\rm Fe}}\,
\frac{\left(^{60}{\rm Fe}/^{56}{\rm Fe}\right)_{t_0,{\rm CHUR}}}{\left(^{60}{\rm Fe}/^{56}{\rm Fe}\right)_{\rm ccSN/AGB}}e^{\lambda_{60}\Delta t}\times 10^4.
\end{equation}
Similarly to Eq. \ref{eq:het}, the isotopic composition of a reservoir formed with an initial $\left({\rm ^{60}Fe/^{56}Fe}\right)_{t_0,r}$ ratio different from CHUR is,
 \begin{equation}
\epsilon ^i{\rm Fe}_r\simeq\frac{\rho ^i_{\rm Fe}-\mu ^i_{\rm Fe}\,\rho ^{57}_{\rm Fe}}{1+\rho ^{56}_{\rm Fe}}\,
\frac{\left(^{60}{\rm Fe}/^{56}{\rm Fe}\right)_{t_0,r}-\left(^{60}{\rm Fe}/^{56}{\rm Fe}\right)_{t_0,{\rm CHUR}}}{\left(^{60}{\rm Fe}/^{56}{\rm Fe}\right)_{\rm ccSN/AGB}}e^{\lambda_{60}\Delta t}\times 10^4.
\label{eq:het2}
\end{equation}
This equation can be used to quantify the degree of heterogeneity in ${\rm ^{60}Fe/^{56}Fe}$ ratios based on $\epsilon ^{58}{\rm Fe}$ measurements.

\section{Derivation of Eq. 3}

The derivation of Eq. 3 is in many respects similar to that presented for Eq. 2. An important difference lies in the fact that Fe and Ni have different chemical behaviors and can therefore be decoupled during or after injection. In the following, we shall assume that Fe and Ni are not fractionated. Let us define $z=^{58}{\rm Ni}_{\rm ccSN/AGB}/^{58}{\rm Ni}_{\rm CHUR}$. This can be rewritten in the form,
\begin{equation}
z=\frac{\rm \left(^{58}Ni/^{54}Fe\right)_{\rm ccSN/AGB}}{\rm \left(^{58}Ni/^{54}Fe\right)_{\rm CHUR}}\frac{\rm ^{54}Fe_{ccSN/AGB}}{\rm ^{54}Fe_{CHUR}}.
\end{equation}
If we now introduce $c=\left(^{58}{\rm Ni}/^{54}{\rm Fe}\right)_{\rm ccSN/AGB}/\left(^{58}{\rm Ni}/^{54}{\rm Fe}\right)_{\rm CHUR}$, it follows,
\begin{equation}
z=\frac{cx}{1+\rho ^{56}{\rm Fe}}.
\end{equation}
For Ni, we can derive an equation very similar to Eq. \ref{eq:ep2} with $z$ in place of $y$,
\begin{equation}
\epsilon ^i{\rm Ni}_0=-\frac{z}{1-z}\left(\rho ^i_{\rm Ni}-\mu ^i_{\rm Ni}\rho ^{62}_{\rm Ni}\right)\times 10^4,
\label{eq:ep3}
\end{equation}
where $\mu ^i_{\rm Ni}=(i-58)/(62-58)$ and $\rho ^i_{\rm Ni}=\left(^i{\rm Ni}/^{58}{\rm Ni}\right)_{\rm ccSN/AGB}/\left(^i{\rm Ni}/^{58}{\rm Ni}\right)_{\rm CHUR}-1$. As discussed before, $cx\ll 1+\rho ^{56}_{\rm Fe}$. For that reason, the expression $z/(1-z)=cx/(1+\rho ^{56}_{\rm Fe}-cx)$ can be approximated by $cx/(1+\rho ^{56}_{\rm Fe})$. We therefore have (Eq. \ref{eq:l3}),
\begin{equation}
\epsilon ^i{\rm Ni}_0\simeq -c\,\frac{\rho ^i_{\rm Ni}-\mu ^i_{\rm Ni}\,\rho ^{62}_{\rm Ni}}{1+\rho ^{56}_{\rm Fe}}\,
\frac{\left(^{60}{\rm Fe}/^{56}{\rm Fe}\right)_{t_0,{\rm CHUR}}}{\left(^{60}{\rm Fe}/^{56}{\rm Fe}\right)_{\rm ccSN/AGB}}e^{\lambda_{60}\Delta t}\times 10^4.
\end{equation}
Note that for $i=60$, the nucleosynthetic anomalies would superimpose on ${\rm ^{60}Ni}$ deficits from lack of ${\rm ^{60}Fe}$ decay (Eq. \ref{eq:l1}). Similarly to Eq. \ref{eq:het}, the isotopic composition of a reservoir formed with an initial $\left({\rm ^{60}Fe/^{56}Fe}\right)_{t_0,r}$ different from CHUR is,
\begin{equation}
\epsilon ^i{\rm Ni}_r\simeq c\,\frac{\rho ^i_{\rm Ni}-\mu ^i_{\rm Ni}\,\rho ^{62}_{\rm Ni}}{1+\rho ^{56}_{\rm Fe}}\,
\frac{\left(^{60}{\rm Fe}/^{56}{\rm Fe}\right)_{t_0,r}-\left(^{60}{\rm Fe}/^{56}{\rm Fe}\right)_{t_0,{\rm CHUR}}}{\left(^{60}{\rm Fe}/^{56}{\rm Fe}\right)_{\rm ccSN/AGB}}e^{\lambda_{60}\Delta t}\times 10^4.
\label{eq:het3}
\end{equation}
This equation can be used to quantify the degree of heterogeneity in ${\rm ^{60}Fe/^{56}Fe}$ ratios based on $\epsilon ^{64}{\rm Ni}$ measurements.

\section{Derivation of Eq. 4}

This equation was derived in another context by \citet{jacobsen84}. Let us consider a reservoir $r$ that evolved with chondritic composition until time $t_i$, at which point Fe/Ni fractionation occured. The present inventory of ${\rm ^{60}Ni}$ is,
\begin{equation}
^{60}{\rm Ni}_r=^{60}{\rm Ni}_{t_i,r}+^{60}{\rm Fe}_{t_i,r}.
\end{equation}
This can be rewritten as,
\begin{equation}
\left({\rm ^{60}Ni}/{\rm ^{58}Ni}\right)_r=\left({\rm ^{60}Ni}/{\rm ^{58}Ni}\right)_{t_i,r}+\left({\rm ^{60}Fe}/{\rm ^{56}Fe}\right)_{t_i}\left({\rm ^{56}Fe}/{\rm ^{58}Ni}\right)_r.
\end{equation}
We note that because the system evolved with chondritic composition until $t_i$, we have $(^{60}{\rm Ni}/^{58}{\rm Ni})_{t_i,r}=(^{60}{\rm Ni}/^{58}{\rm Ni})_{t_i, {\rm CHUR}}$,
\begin{equation}
\left({\rm ^{60}Ni}/{\rm ^{58}Ni}\right)_r=\left({\rm ^{60}Ni}/{\rm ^{58}Ni}\right)_{t_i,{\rm CHUR}}+\left({\rm ^{60}Fe}/{\rm ^{56}Fe}\right)_{t_i}\left({\rm ^{56}Fe}/{\rm ^{58}Ni}\right)_r.
\end{equation}
The same equation can be posed for CHUR,
\begin{equation}
\left({\rm ^{60}Ni}/{\rm ^{58}Ni}\right)_{\rm CHUR}=\left({\rm ^{60}Ni}/{\rm ^{58}Ni}\right)_{t_i,{\rm CHUR}}+\left({\rm ^{60}Fe}/{\rm ^{56}Fe}\right)_{t_i}\left({\rm ^{56}Fe}{\rm ^{58}Ni}\right)_{\rm CHUR}.
\end{equation}
Forming the difference between the last two equation,
\begin{equation}
\left({\rm ^{60}Ni}/{\rm ^{58}Ni}\right)_r-\left({\rm ^{60}Ni}/{\rm ^{58}Ni}\right)_{\rm CHUR}=\left({\rm ^{60}Fe}/{\rm ^{56}Fe}\right)_{t_i}\left[\left({\rm ^{56}Fe}/{\rm ^{58}Ni}\right)_r-\left({\rm ^{56}Fe}/{\rm ^{58}Ni}\right)_{\rm CHUR}\right].
\end{equation}
Dividing the two sides by $\left(^{60}{\rm Ni}/^{58}{\rm Ni}\right)_{\rm CHUR}$, it is easy to rearrange the equation and show that (Eq. \ref{eq:l4}),
\begin{equation}
\epsilon ^{60}{\rm Ni}=Q\left(\frac{\rm ^{60}Fe}{\rm ^{56}Fe}\right)_{t_i}f_{\rm Fe/Ni},
\end{equation}
where $Q=\left(^{56}{\rm Fe}/^{60}{\rm Ni}\right)_{\rm CHUR}\times 10^4$ and $f_{\rm Fe/Ni}={\rm (Fe/Ni)_r/(Fe/Ni)_{CHUR}}-1$.




\clearpage

\renewcommand{\thefigure}{\arabic{figure}}
\setcounter{figure}{0}
\renewcommand{\thetable}{\arabic{table}}
\setcounter{table}{0}

\begin{figure}
\epsscale{0.5}
\plotone{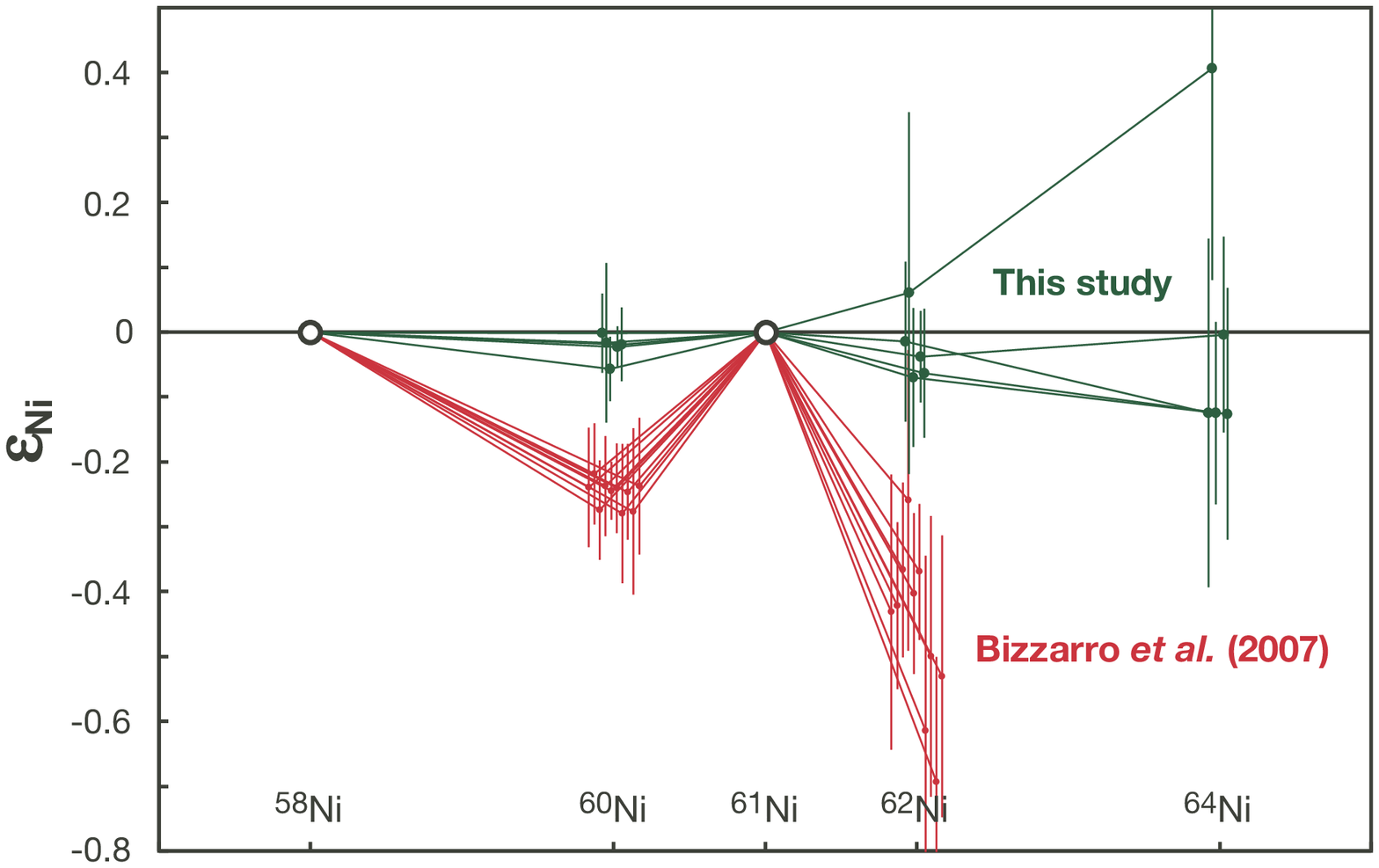}
\caption{Comparison between Ni data from this study (Table \ref{table1}) and \citet{bizzarro07} for differentiated meteorites. Identical notation \citep[$\epsilon ^i{\rm Ni}/^{58}{\rm Ni}$,]{birck88,bizzarro07} and correction for mass-dependent fractionation \citep[internal normalization to a fixed ${\rm ^{61}Ni/^{58}Ni}$ ratio of 0.0167442 using the exponential law,][]{gramlich89,marechal99} were used. Because of its low abundance (0.91 atom \%) and the presence of a major ${\rm ^{64}Zn}$ interference (48.63 atom \%), ${\rm ^{64}Ni}$ data were not reported in \citet{bizzarro07}. [{\it See the electronic edition of the Journal for a color version of this figure}].
\label{fig:compbiz}
}
\end{figure}

\clearpage

\begin{figure}
\epsscale{0.5}
\plotone{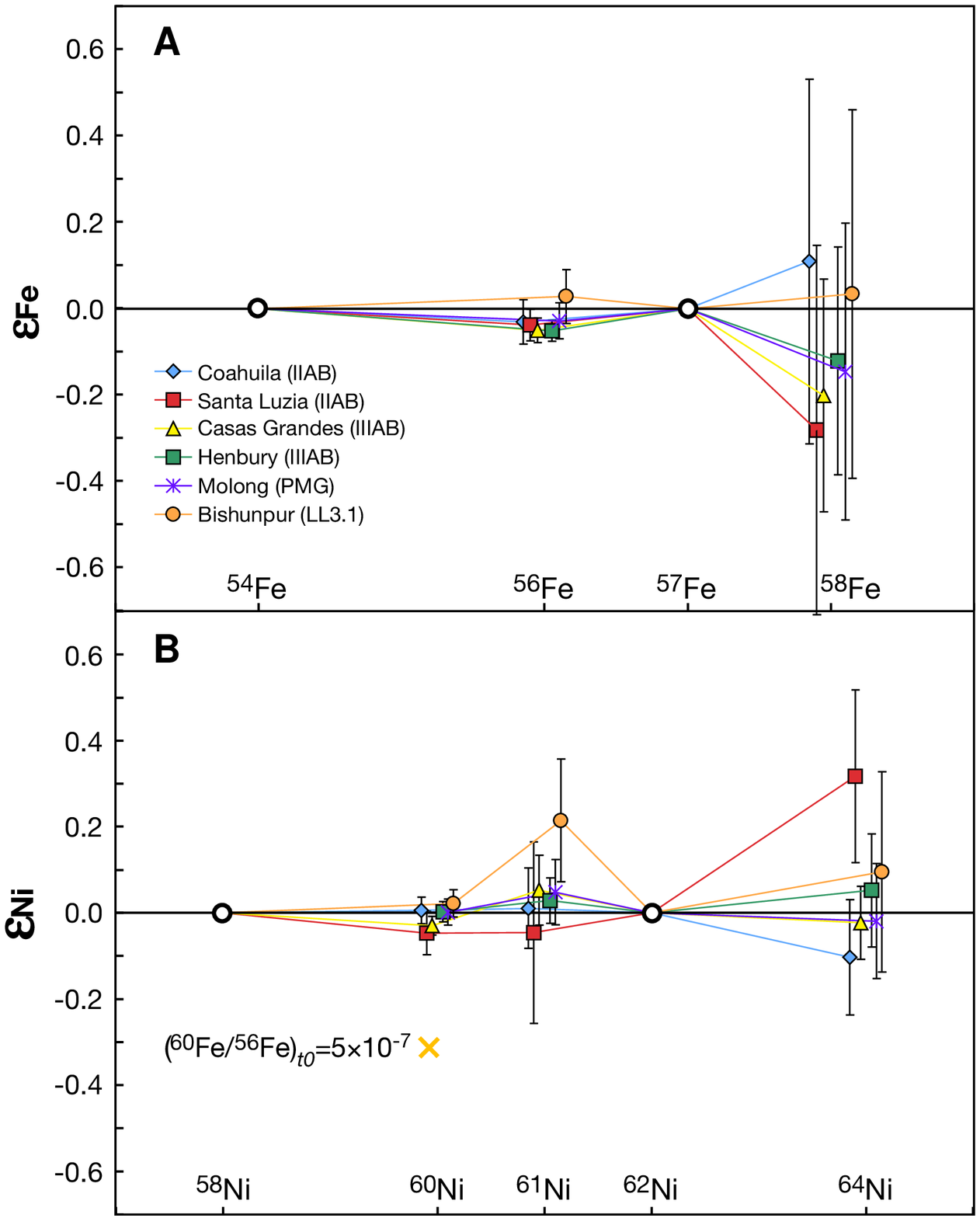}
\caption{Fe and Ni isotope data for meteoritic metal (Table \ref{table1}). The compositions are in $\epsilon$ units corrected for mass-dependent fractionation by fixing the ratios ${\rm ^{57}Fe/^{54}Fe}=0.3625663$ \citep{taylor92} and ${\rm ^{62}Ni/^{58}Ni}=0.0533886$ \citep{gramlich89} using the exponential law \citep{marechal99}. Fe has normal (terrestrial) isotopic composition (A). If iron meteorites did not incorporate ${\rm ^{60}Fe}$ and the initial $\left(^{60}{\rm Fe}/^{56}{\rm Fe}\right)_{t_0,{\rm CHUR}}$ ratio was $\sim 5\times 10^{-7}$ \citep{tachibana06}, then one would expect to find $\epsilon ^{60}{\rm Ni}_0=-10^4\left(^{60}{\rm Fe}/^{56}{\rm Fe}\right)_{t_0,{\rm CHUR}}\left(^{56}{\rm Fe}/^{60}{\rm Ni}\right)_{\rm CHUR}\simeq -0.31$ in iron meteorites (large cross) relative to chondrites and Earth, which is not observed (B). This conclusion does not depend on the pair of isotopes (${\rm ^{61}Ni/^{58}Ni}$ or ${\rm ^{62}Ni/^{58}Ni}$) that is used for correcting the measurements for mass-dependent fractionation (Table \ref{table1}, Fig. \ref{fig:compbiz}). [{\it See the electronic edition of the Journal for a color version of this figure}].
\label{fig:compilcomp}
}
\end{figure}

\clearpage

\begin{figure}
\epsscale{0.5}
\plotone{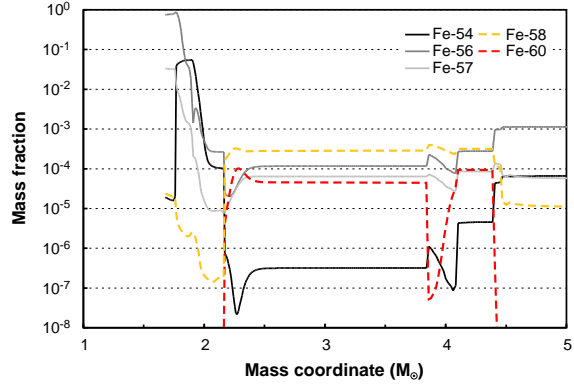}
\caption{Post-supernova profile of Fe isotope abundances as a function of mass coordinate for a 19 M$_{\odot}$ cc-SN progenitor computed from zonal yields 25,000 s after core bounce \citep[see][for details]{rauscher02}. While ${\rm ^{54}Fe}$, ${\rm ^{56}Fe}$, and ${\rm ^{57}Fe}$ are produced as radioactive progenitors in the internal regions of the cc-SN by nuclear statistical equilibrium, the neutron-rich isotopes ${\rm ^{58}Fe}$ and ${\rm ^{60}Fe}$ are produced in more external regions by neutron-capture reactions on pre-existing Fe isotopes. [{\it See the electronic edition of the Journal for a color version of this figure}].
\label{fig:tommy}}
\end{figure}

\clearpage

\begin{figure}
\epsscale{0.5}
\plotone{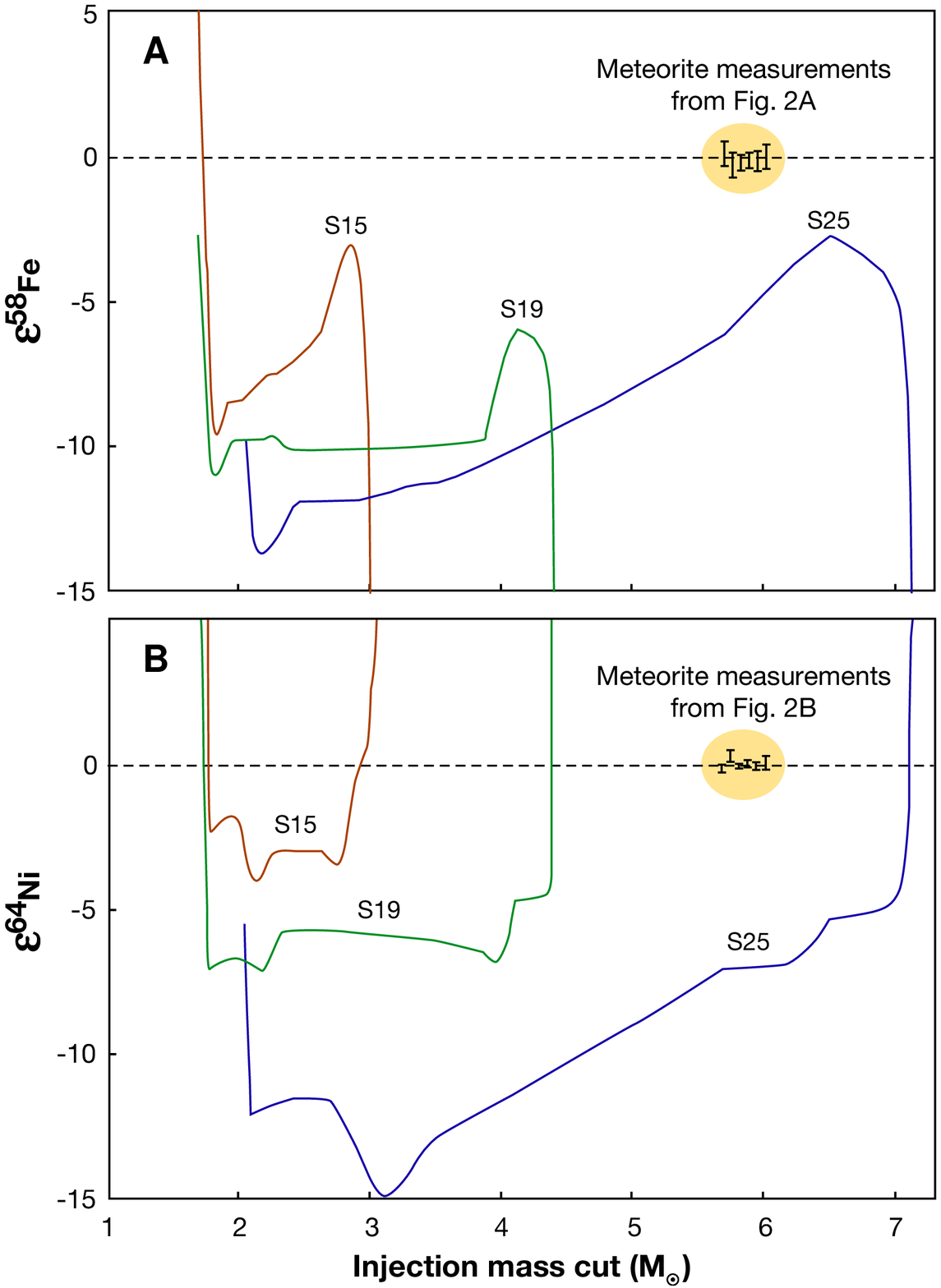}
\caption{Comparison between measured Fe and Ni isotopic compositions in meteoritic metal (Table \ref{table1}) with predicted collateral isotopic effects if ${\rm ^{60}Fe}$ from a cc-SN was heterogeneously distributed in the ESS (Eqs. 2, 3). We used Eqs. 2 and 3 with ${\rm (^{60}Fe/^{56}Fe)_{t_0,{\rm CHUR}}=5\times 10^{-7}}$ \citep{tachibana06}, $\Delta t=0$, and the yields for 15, 19, and 25 M$_{\odot}$ cc-SN progenitors \citep{rauscher02} to compute ${\rm \epsilon ^{58}{Fe}}$ and ${\rm \epsilon ^{64}{Ni}}$ anomalies as a function of injection mass cut \citep[mass coordinate above which matter from the cc-SN ejecta is injected into the solar system,][]{meyer00}. The 20 and 21 M$_\odot$ models of \citet{rauscher02} are not plotted because S21 is redundant with S19, and S20 exhibits a singular and possibly unrealistic convection pattern \citep[merging of O-, Ne-, and C-burning shells,][]{rauscher02,tur07}. If portions of the solar system did not incorporate ${\rm ^{60}Fe}$, one would expect to find anomalous $\epsilon ^{58}{\rm Fe}$ (A) and $\epsilon ^{64}{\rm Ni}$ (B), which is not observed. For each cc-SN model, there is a certain injection mass cut above which the amount of ${\rm ^{60}Fe}$ ejected decreases, requiring unrealistic dilution factors for explaining the ${\rm ^{60}Fe/^{56}Fe}$ ratio measured in meteorites (Eq. \ref{eq:x1}). This is why for each model, the results are not plotted above a certain injection mass cut. [{\it See the electronic edition of the Journal for a color version of this figure}].
\label{fig:modelres}}
\end{figure}

\clearpage

\begin{figure}
\epsscale{0.5}
\plotone{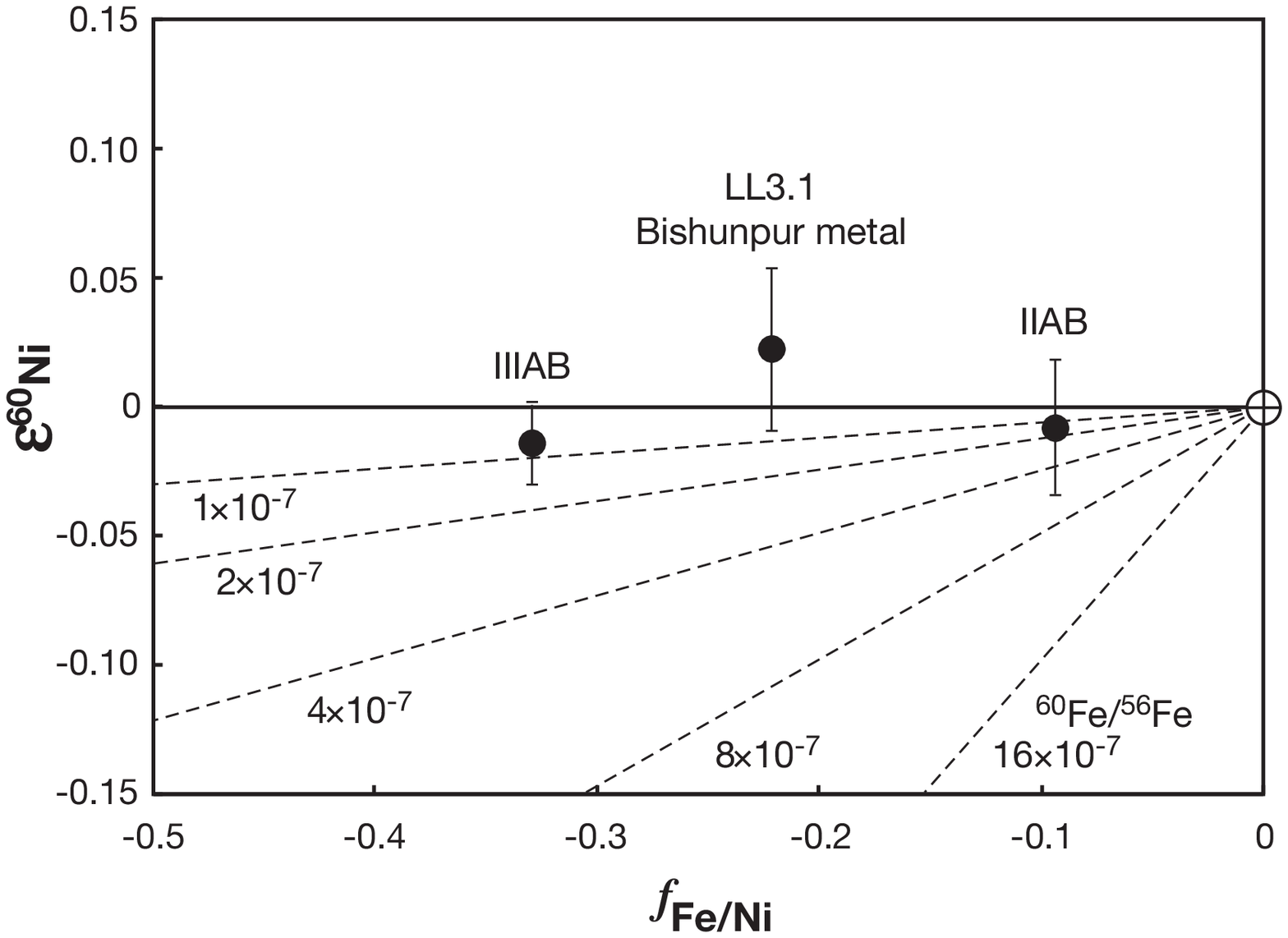}
\caption{ Isochron diagram between $\epsilon ^{60}{\rm Ni}$ (Table \ref{table1}) and $f_{\rm Fe/Ni}$ (Eq. 4). For Bishunpur metal, $f_{\rm Fe/Ni}$ was calculated using the ${\rm ^{56}Fe/^{58}Ni}$ ratio reported in Table 2 of \citet{cook06}. For IIAB and IIIAB iron meteorites, the $\epsilon ^{60}{\rm Ni}$ are the weighted averages of the values for each group (Table \ref{table1}) and the $f_{\rm Fe/Ni}$ values were calculated using the Fe/Ni ratios of parental melts \citep{jones83}. The dashed lines correspond to expected correlations for different values of the ${\rm ^{60}Fe/^{56}Fe}$ ratio.
\label{fig:isochron}
}
\end{figure}

\clearpage

\begin{table}
{\scriptsize
\begin{center}
\caption{Iron and nickel isotopic compositions of meteoritic metal.}
\begin{tabular}{lcccccc}
\tableline\tableline
\multicolumn{7}{c}{$\epsilon ^{i}{\rm Ni}/^{58}{\rm Ni}$ normalized to ${\rm ^{62}Ni/^{58}Ni=0.0533886}$}\\
Sample (type) & \# & $\epsilon^{58}_{\rm Ni}$ & $\epsilon^{60}_{\rm Ni}$ & $\epsilon^{61}_{\rm Ni}$ & $\epsilon^{62}_{\rm Ni}$ & $\epsilon^{64}_{\rm Ni}$\\
Coahuila (IIAB) & 15 &       0 & $0.006\pm 0.031$  & $0.011\pm 0.093$  & 0 & $-0.103\pm 0.134$\\
Santa Luzia (IIAB) & 8 &     0 & $-0.047\pm 0.050$ & $-0.046\pm 0.211$ & 0 & $0.317\pm 0.200$\\
Casas Grandes (IIIAB) & 30 & 0 & $-0.030\pm 0.022$ & $0.053\pm 0.081$  & 0 & $-0.023\pm 0.085$\\
Henbury (IIIAB) & 17 &       0 & $0.003\pm 0.023$  & $0.029\pm 0.053$  & 0 & $0.053\pm 0.131$\\
Molong (PMG) & 33 &          0 & $0.002\pm 0.031$  & $0.048\pm 0.076$  & 0 & $-0.019\pm 0.134$\\
Bishunpur (LL3.1) & 14 &     0 & $0.022\pm 0.031$  & $0.215\pm 0.143$  & 0 & $0.095\pm 0.233$\\
\\
\multicolumn{7}{c}{$\epsilon ^{i}{\rm Ni}/^{58}{\rm Ni}$ normalized to ${\rm ^{61}Ni/^{58}Ni=0.0167442}$}\\
Sample (type) & \# & $\epsilon^{58}_{\rm Ni}$ & $\epsilon^{60}_{\rm Ni}$ & $\epsilon^{61}_{\rm Ni}$ & $\epsilon^{62}_{\rm Ni}$ & $\epsilon^{64}_{\rm Ni}$\\
Coahuila (IIAB) & 15 &       0 & $-0.001\pm 0.061$  & 0 & $-0.014\pm 0.123$ & $-0.124\pm 0.269$\\
Santa Luzia (IIAB) & 8 &     0 & $-0.016\pm 0.123$  & 0 & $0.061\pm 0.279$  & $0.407\pm 0.325$\\
Casas Grandes (IIIAB) & 30 & 0 & $-0.057\pm 0.050$  & 0 & $-0.070\pm 0.107$ & $-0.125\pm 0.141$\\
Henbury (IIIAB) & 17 &       0 & $-0.023\pm 0.032$  & 0 & $-0.038\pm 0.070$ & $-0.003\pm 0.151$\\
Molong (PMG) & 33 &          0 & $-0.019\pm 0.057 $ & 0 & $-0.063\pm 0.100$ & $-0.126\pm 0.194$\\
Bishunpur (LL3.1) & 14 &     0 & $-0.121\pm 0.092$  & 0 & $-0.284\pm 0.189$ & $-0.324\pm 0.287$\\
\\
\multicolumn{7}{c}{$\epsilon ^{i}{\rm Fe}/^{54}{\rm Fe}$ normalized to ${\rm ^{57}Fe/^{54}Fe=0.3625663}$}\\
Sample (type) & \# & $\epsilon^{54}_{\rm Fe}$ & $\epsilon^{56}_{\rm Fe}$ & $\epsilon^{57}_{\rm Fe}$ & $\epsilon^{58}_{\rm Fe}$ & \\
Coahuila (IIAB) & 12 &       0 & $-0.031\pm 0.051$ & 0 & $0.109\pm 0.422$ &\\
Santa Luzia (IIAB) & 12 &    0 & $-0.038\pm 0.037$ & 0 & $-0.282\pm 0.427$ &\\
Casas Grandes (IIIAB) & 24 & 0 & $-0.051\pm 0.028$ & 0 & $-0.202\pm 0.270$ &\\
Henbury (IIIAB) & 24 &       0 & $-0.052\pm 0.024$ & 0 & $-0.122\pm 0.264$ &\\
Molong (PMG) & 12 &          0 & $-0.029\pm 0.042$ & 0 & $-0.146\pm 0.344$ &\\ 
Bishunpur (LL3.1) & 12 &     0 & $0.028\pm 0.063$  & 0 & $0.033\pm 0.427$ &\\
\tableline
\end{tabular}
\tablenotetext{}{Uncertainties are 95 \% confidence intervals.}
\label{table1}
\end{center}
}
\end{table}

\end{document}